\PassOptionsToPackage{table,xcdraw}{xcolor}

\documentclass{article}
\usepackage[T1]{fontenc} 
\usepackage[utf8]{inputenc} 
\usepackage{ismir,amsmath,cite,url}
\usepackage{graphicx}
\usepackage{color}
\usepackage{multirow} 
\usepackage{xcolor}
\usepackage{amsfonts}

\usepackage{acronym}
\acrodef{uamat}[\texttt{auSP}]{Audio+User based Situation Prediction}
\acrodef{sp}[\texttt{duSP}]{Device+User based Situation Prediction}
\acrodef{fc}[FC]{fully connected}

\usepackage[htt]{hyphenat}

\title{Exploiting device and audio data to tag music with user-aware listening contexts}

\multauthor
{Karim M. Ibrahim$^{1,2}$ \hspace{1cm} 
Elena V. Epure$^2$ \hspace{1cm} 
Geoffroy Peeters$^1$ \hspace{1cm} 
Ga{\"e}l Richard$^1$}
{ $^1$ LTCI, T\'el\'ecom Paris, Institut Polytechnique de Paris\\
$^2$ Deezer Research\\
{\tt\small karim.ibrahim@telecom-paris.fr}
}
\def\authorname{Karim M. Ibrahim, Elena V. Epure, Geoffroy Peeters, Ga{\"e}l Richard}

\begin{document}

\maketitle

\begin{abstract}
As music has become more available especially on music streaming platforms, people have started to have distinct preferences to fit to their varying listening situations, also known as context. Hence, there has been a growing interest in considering the user's situation when recommending music to users. Previous works have proposed user-aware autotaggers to infer situation-related tags from music content and user's global listening preferences. However, in a practical music retrieval system, the autotagger could be only used by assuming that the context class is explicitly provided by the user. In this work, for designing a fully automatised music retrieval system, we propose to disambiguate the user's listening information from their stream data. Namely, we propose a system which can generate a situational playlist for a user at a certain time 1) by leveraging user-aware music autotaggers, and 2) by automatically inferring the user's situation from stream data (e.g. device, network) and user's general profile information (e.g. age). Experiments show that such a context-aware personalized music retrieval system is feasible, but the performance decreases in the case of new users, new tracks or when the number of context classes increases. 
\end{abstract}

\section{Introduction} \label{sec:introduction}

Since the invention of recorded music, people have been shifting from consuming music as a main activity in a live setting, to using music as a background activity as they go through the day. 
With the growing availability of music on streaming platforms, people developed distinct preferences for the varying listening situations, also known as context \cite{north1996situational}.
Consequently, there has been a growing interest in considering the user's situation when automatically recommending music to users.

Previous works have proposed user-aware autotaggers to infer situation-related tags from music content and user's global listening preferences \cite{ibrahim2020should}. However, in a practical music retrieval system, the autotagger could be only used by assuming that the context class is explicitly provided by the user.
In this work, we perform a study to evaluate the feasibility of inferring the listening situation.
The listening situation for our system is an activity, location, or time that is influencing the listener's preferences. 

\begin{figure}
  \centering
  \includegraphics[width=1\linewidth]{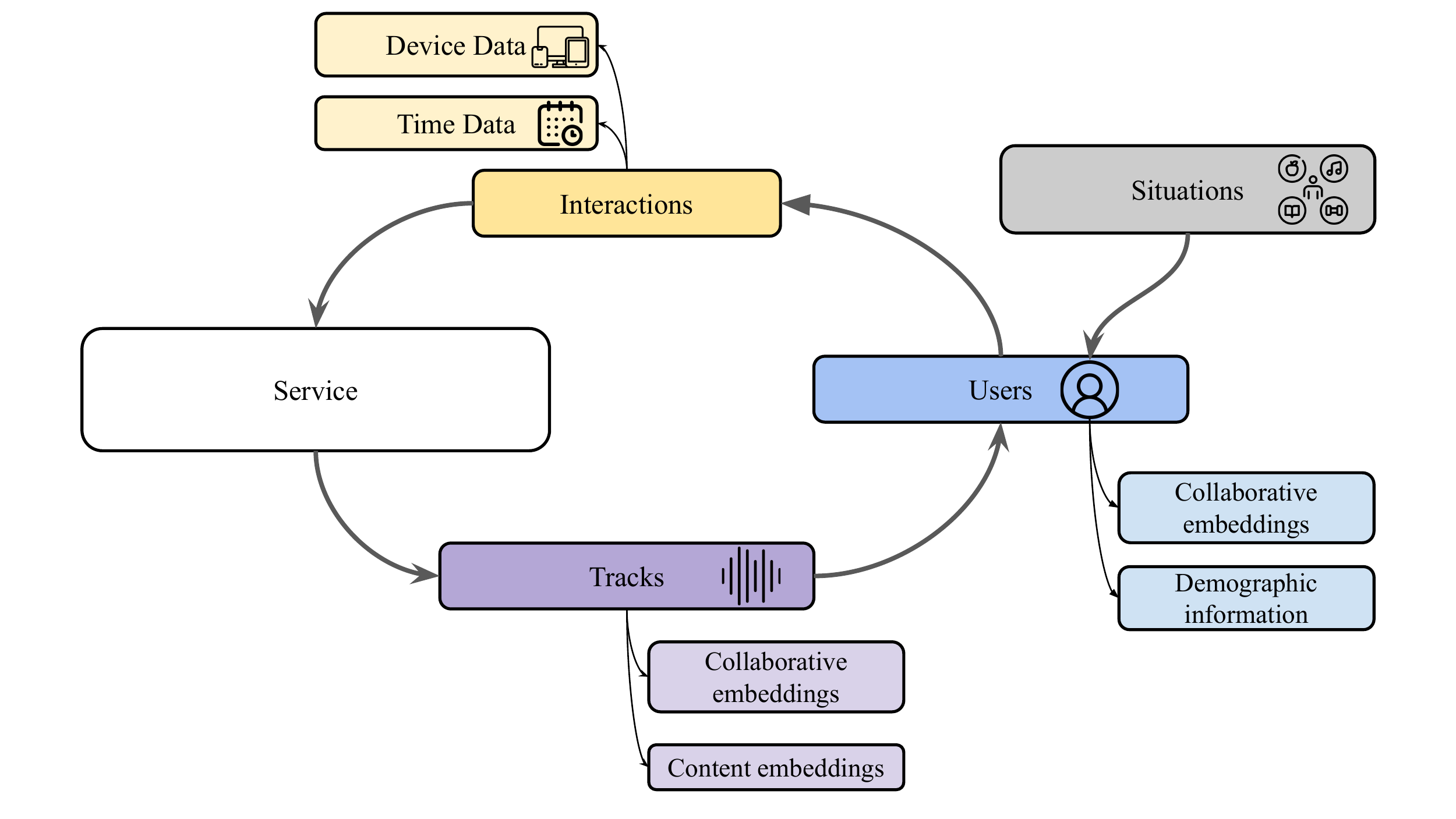}
  \caption{The available data to online music streaming services.} 
  \label{fig:process}
\end{figure}

The process of music streaming from the perspective of our proposed approach can be found in Figure \ref{fig:process}. 
We find that the music service is informed of the users, their track history, plus their past and current interactions with the service, i.e. the device and time data sent during an active session. However, the service is unaware of the influencing listening situation. 
Our goal is to utilize the available information for the service to infer the listening situation and the suitable tracks for the inferred situation. 
We propose an approach that infers the potential context from the user interactions in near real time, while the tagging of tracks with their potential listening situation happens in the background using autotaggers. Both systems are user-aware.

Our contributions in this paper are:
1) a large dataset of tracks, device data, and user embeddings labeled with their situational use through a rigorous labelling pipeline;
2) an extended evaluation of music autotaggers in predicting personalized situational tags in various scenarios; 
3) a simple, yet effective model that ranks the potential listening situations for a given user based on the transmitted data from the device to the service. 


\section{Related work}

 Our proposed approach is related to two different problems: music autotagging with contextual tags, and instant prediction of the user's listening situation. Previous work has already showed that listening situation (i.e. context) has a strong 
 influence on the user's preferences \cite{north1996situational,greasley2011exploring, gorgoglione2011effect,hansen2020contextual}. Hence, context has become an important factor for reaching a personalized user experience \cite{kaminskas2012contextual}. 
 
On one hand, music content is highly complex and is often challenging to be analyzed and described in human readable terms.  This missing link between the content of the music and a set of semantic descriptors is referred to as the ``semantic gap'' \cite{celma2008foafing}. One common way, which is often used when searching for or organising music, is the intended listening situation \cite{ourpaper}. Unlike most tags that depend solely on the music content, certain tags depend also on the user \cite{schedl2013neglected, korzeniowski2020mood}. There has been a recent work on predicting personalized situation-related tags from music content and user embeddings \cite{ibrahim2020should}, which we adopt here too. 

 On the other hand, the listening situation, e.g. activity or location, can change frequently, which leads to changes in user preferences.
 Explicitly inferring the listening situation is a challenging task that has only been studied on a small scale \cite{wang2012context}. 
 We aim at addressing this missing link by performing an extensive study on predicting the listening situation using available device data. In order to employ the personalized autotagging approach in an actual real-world setting, it is also important to be able to predict when a specific listening situation is being experienced.

\section{Objective and Proposed Approach}
\label{method}

A \textit{session} consists of a sequence of audio-tracks $a$ a given user $u$ is listening to over time $t$ on a music streaming service in a continuous time span\footnote{without any break longer than a pre-defined gap, defined here as 20 minutes as proposed by \cite{hansen2020contextual}}.
A session is therefore defined as a sequence of \textit{streams} which are each a tuple (audio-track $a$, user $u$, device data $\mathbf{d}_u^{(t)}$).

A \textit{situational (or contextual) session} is a session resulting from listening to tracks in a certain situation (or context) $c$ such as ``gym''.
However, in our case, in order to gather a ground-truth dataset, we consider that a situational session can also result from listening to a playlist that contains a context-related keyword in the title\footnote{
The underlying assumption is that if the user started streaming a playlist with a certain title related to a situation (or context), most likely the intention of the user was to play something suitable for that situation (or context) \cite{playlistName}.}.
A situational (or contextual) session is defined as a sequence of tuples (audio-track $a$, user $u$, device data $\mathbf{d}_u^{(t)}$, situation $c$).

Our objectives is to propose for a given user $u$ a session (sequence of audio-tracks $a$) that fits their current situation $c$.
However, since we don't know its current situation $c$ we estimate it based on its device data $\mathbf{d}_u^{(t)}$ (such as the time of the day, day of the week, or type of network connections).

\subsection{Proposed approach}

To do so, we first estimate for each pair audio-track/user $(a,u)$ its situation $\hat{c}_{a,u}$. 
In other words, we estimate in which situation $c$ the user $u$ would intend to use the track $a$.
This is done using an \ac{uamat} trained to estimate situation tags $c$ given as input a pair (audio-track $a$, user $u$).
This is done offline on the server side and stored in a database.

We then estimate in real-time (with a lightweight model on the client side) for a given user $u$ and the transmitted data from its device to the service $\mathbf{d}_u^{(t)}$, its potential current situation $\hat{c}_{d,u}$.
This is done using a \ac{sp} trained to estimate situation tags $c$ given as input a pair (device-data $\mathbf{d}_u^{(t)}$, user $u$).
\ac{sp} provides us with a list of the most likely situations $\hat{c}_{d,u}$ (ranked from the most to the less likely).

Finally, to create the situational playlist, we simply select the audio tracks $a$ for which situation $\hat{c}_{a,u}$ matches the most-likely current situations of the user $\hat{c}_{d,u}$.

\begin{figure} 
  \centering
  \includegraphics[width=\linewidth]{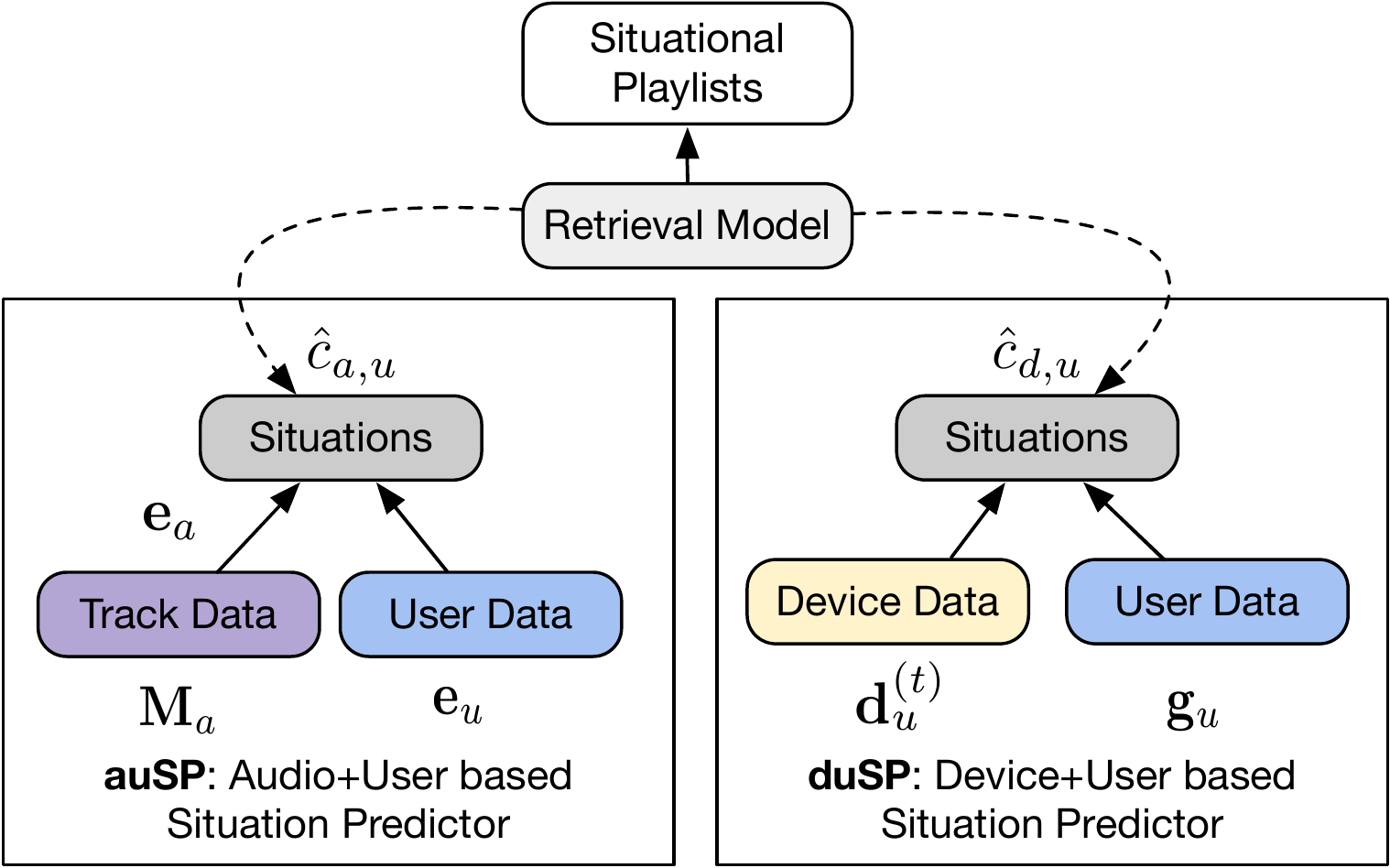}
  \caption{Overview of the system to generate a situational playlist. 
  The left side (\ac{uamat}) tags each track/user pair with a situational tag. 
  The right side (\ac{sp}) ranks the potential situations for a device/user pair to be presented to the user.}
  \label{fig:system}
\end{figure}

Figure~\ref{fig:system} indicates the overall architecture.

\subsection{Data description}

We first describe what are exactly the data for the tracks $a$, the users $u$ and the devices.

\textbf{Track data $\mathbf{M}_a$.} 
For each audio-track $a$, we retrieve its 30~s. snippet from the Deezer API.
We represent $a$ by its 96 Mel-bands $\times$ 646 frames matrix $\mathbf{M}_a$.

\textbf{User data $\mathbf{e_u}$ and $\mathbf{g}_u$.}
Representing the users can be achieved through various versatile techniques. 
Consistent with our requirements (lightweight model and preserving privacy), we choose to represent the users using the basic data available during streaming. 
We use two different representations of the user that will be used for estimating $\hat{c}_{a,u}$ and $\hat{c}_{d,u}$ respectively.

For the \ac{uamat} (estimation of $\hat{c}_{a,u}$) we use a user embedding $\mathbf{e_u}$. 
Similar to previous works on \ac{uamat}, we used the users' listening history to derive user embeddings that encode their listening preferences. 
We compute these embeddings through matrix factorization of the user/track interactions matrix, leading to a 128-d embedding vector per user, which in commonly used to generate representations \cite{lee1999learning}. 
The constructed matrix uses all the tracks available in the catalogue to model the user preferences, i.e. it is not computed exclusively with the tracks included in our dataset. 
The computed embeddings will be published with the dataset for reproducibility.

For the \ac{sp} (estimation of $\hat{c}_{d,u}$), we use the basic demographic data $\mathbf{g}_u$ of the user recorded during registration. 
This data is composed of: \texttt{|age,country,gender|}. While this data selection is prone to errors and short of fully representing the users, it is consistent with our requirements of using basic always-available data. 

\textbf{Device data $\mathbf{d}_u^{(t)}$.} 
We collect only basic data sent by the device to the service and selected those that deemed relevant to the situation prediction. 
The data are: the time stamp (in local time), day of the week, device used and network used. 
Additionally, we extend the time/day data with circular representation of the time and day similar to the one used in \cite{herrera2010rocking}. The final feature vector representing device data is made of 8 features: \texttt{linear-time, linear-day, circular-time-X, circular-time-Y, circular-day-X, circular-day-Y, device-type, network-type}. 
The \texttt{device-type} can be: \texttt{mobile}, \texttt{desktop} (e.g. a laptop), or \texttt{tablet}.
The \texttt{network-type} can be: 
\texttt{mobile} (a connection through cellular data),
\texttt{wifi} (a WiFi connection),
\texttt{LAN} (a connection through wired Ethernet), or
\texttt{plane} (an offline stream from a device without a connection). 

\begin{table}
\small
\centering
\caption{Summary of the notations}
\label{notation}
\begin{center}
\begin{tabular}{l l l l}
\hline
Symbol  & Definition   &  Dimension \\ \hline
$a$ & an audio track \\ \hline
$\;\;\;\; \mathbf{M}_a$   & Mel-spectrogram of $a$  & $\mathbb{R}^{96\times 646}$\\ \hline
$\;\;\;\; \mathbf{e}_a$   & Embedding of $a$ & $\mathbb{R}^{256}$\\ \hline
$u$ & a user \\ \hline
$\;\;\;\; \mathbf{e}_u$   & Embedding of $u$ & $\mathbb{R}^{128}$\\ \hline
$\;\;\;\; \mathbf{g}_u$   & Demographics data of $u$ & $\mathbb{R}^{3}$ \\ \hline
$\;\;\;\; \mathbf{d}_u^{(t)}$   & Device data of $u$ at time $t$ & $\mathbb{R}^{8}$\\ \hline
$c$ & a situation (or context) & \\ \hline
$s$ & a stream, a tuple $(a,u,\mathbf{d}_u^{(t)},c)$ \\ \hline
\end{tabular}
\end{center}
\end{table}

\section{Collecting the data}

Pichl et al.~and Ibrahim et al. proposed methods for labelling streaming sessions with a situational tag by leveraging playlist titles \cite{playlistName,ibrahim2020should}.
Although sometimes prone to errors and false positives, playlist titles provide an appropriate proxy for labelling streams with tags  \cite{playlistName,ibrahim2020should}.
Users create playlists with a common theme according to their use \cite{playlistName}. 
One common theme of these playlists is the listening situation. 

First, we collected a set of situational keywords used previously in the literature \cite{north1996situational,wang2012context,gillhofer2015iron}. 
We extended these keywords by adding synonyms and hashtags that are frequently used on Twitter to refer to music listening. 
Afterwards, we retrieve from all public playlists from Deezer\footnote{\url{www.deezer.com}}, an online music streaming service we were given access to, those playlists that include any of the keywords in their ``stemmed'' title. 
We then filtered out playlists that contained more than 100 tracks\footnote{to increase the chance that playlists reflect a selection of situation-related tracks, and not randomly added ones} or where a single artist or album represented more than 25\% of the playlist, similar to \cite{ibrahim2020should}. 

From the extensive list of situational keywords and their corresponding playlists, we settled on three different subsets with an increasing number $C$ of situational tags (4, 8, and 12):
\texttt{work, gym, party, sleep | morning, run, night, dance | car, train, relax, club}.

These tags were selected by popularity\footnote{Estimated as the number of corresponding playlists in the service catalogue.}.
We used these situations as independent tags without attempting to merge potentially similar activities and places (e.g. "party" and "dance"). 
Working with three situational tag sets (of increasing $C$) allowed us to observe how the system performs as the complexity of the problem increases.

We then focused on the users who actively listened to these playlists and retrieve the device data of those users while they were actively listening to the playlist.
This resulted in a set of streams each described by an audio-track $a$, a user $u$, a playlist with a situational keyword $c$ in the title, along with the device data $\mathbf{d}_u^{(t)}$ sent during this stream.
Note that an audio-track/user/device triplet have a joint tag, none of them are tagged individually. 

To ensure high quality data, we selected the month of August 2019 for inspection, because this period had more stable use patterns, before the Covid-19 pandemic. 
We had access to data from two locations: France and Brazil. These two locations were provided because they have the most active users in Deezer, while being in two distinctive time zones and seasons. 
This allowed us to perform our study on diverse data with different sources and patterns. We release the dataset\footnote{\url{https://zenodo.org/record/5552288}} along with the code\footnote{\url{https://github.com/Karimmibrahim/Situational_Session_Generator}}.

\subsection{Dataset Analysis}
\label{analysis}

\begin{figure*}[h]
\minipage{0.31\textwidth}
  \includegraphics[width=1.2\textwidth]{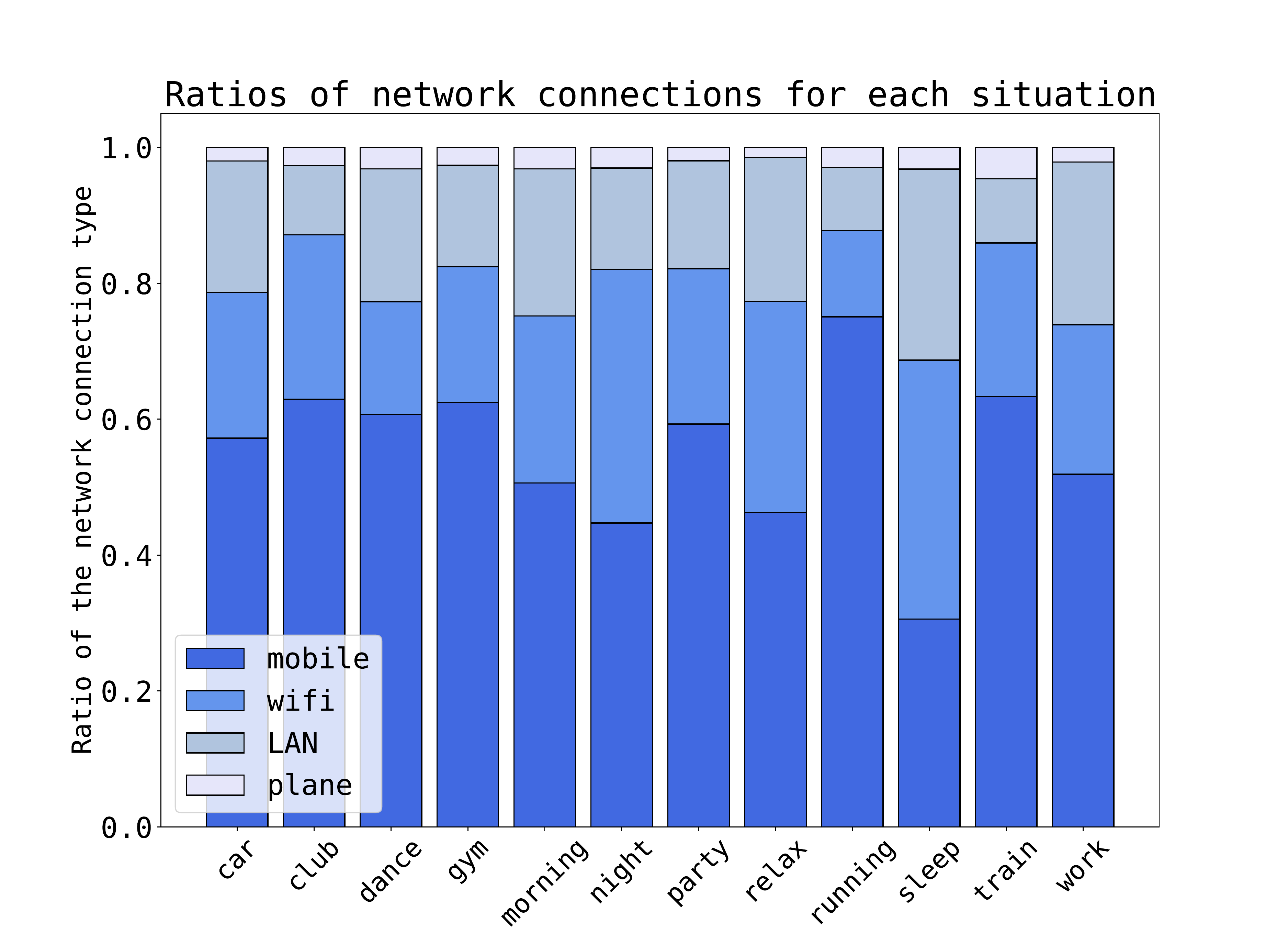}
  \caption{Network across situations $c$}
  \label{fig:network}
\endminipage \quad
\minipage{0.31\textwidth}
  \includegraphics[width=1.2\textwidth]{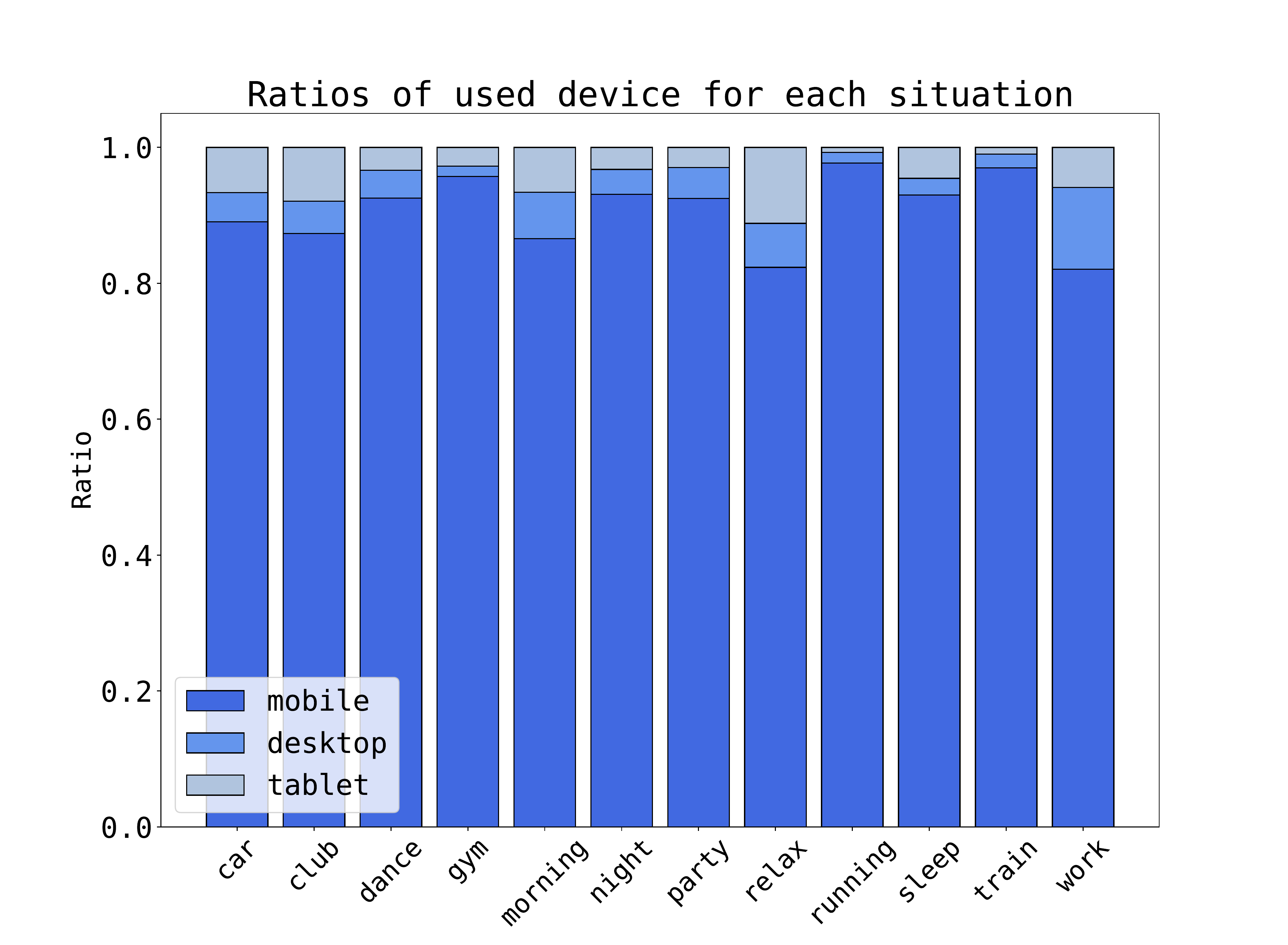}
  \caption{Device across situations $c$}
  \label{fig:device}
\endminipage \quad
\minipage{0.31\textwidth}
      \includegraphics[width=1.2\textwidth]{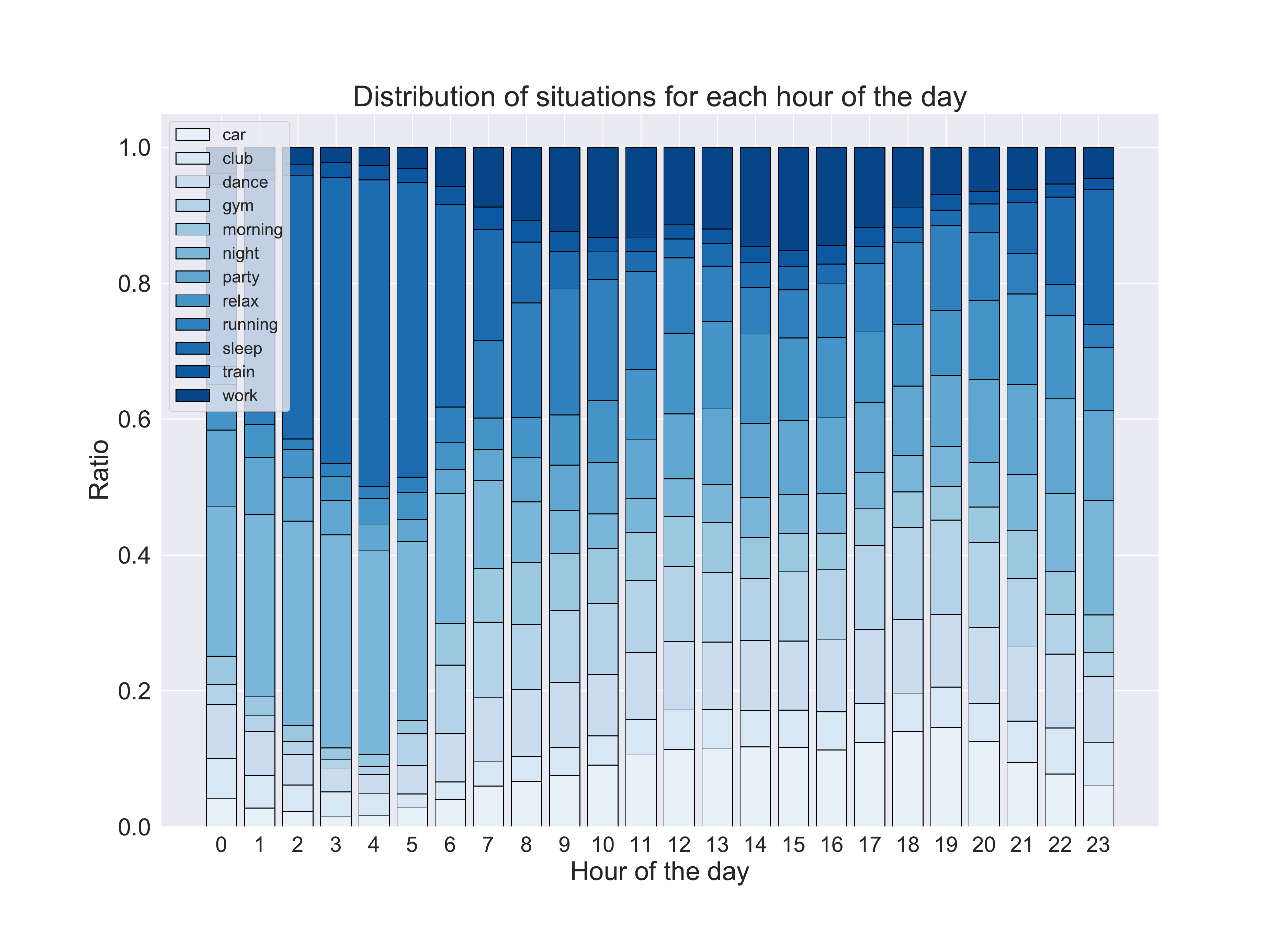}
      \caption{Distributions of situations $c$ over hours of the day}
      \label{fig:hours}
\endminipage \quad
\end{figure*}

As a sanity check on the collected data, we plot the distribution of the situations $c$ across the different device-data.
Figure \ref{fig:network} shows the ratio of the used \texttt{network-type} to connect to the service across situations $c$. 
We observe variations that correspond to what is expected from each situation, i.e. outdoors vs. indoors. 
However, we also find certain networks that do not match the expectations, e.g. LAN connections in a car situation, which represents noise in the dataset that can be a continuation of already existing sessions that moved indoors. 
Figure \ref{fig:device} represents the used \texttt{device-type} across situations $c$. 
We notice that most users overwhelmingly use mobile device in most cases, with small variations that also match expectations of indoor and outdoor situations. 
Finally, Figure \ref{fig:hours} shows the distribution of all situations for each hour of the day. 
Similarly, we find predictable patterns for each situation ranging from night-related situation in the early hours that gradually progress as the time passes. 
These patterns support the hypothesis of using playlist titles as proxy for inferring the actual listening situation.

\section{Detailed models description}

\subsection{\acf{uamat}}
\label{autoDetails}

The \ac{uamat} estimates the probability of each situation $c \in \{1 \ldots C\} $ given a pair (track $a$ represented by $\mathbf{M}_a$, user $u$ represented by $e_u$): $P(c | \mathbf{M}_{a}, \mathbf{e_u})$.
It is implemented as a Deep Neural Network $\hat{c}_{a,u} = f_{\theta}(\mathbf{M}_a, \mathbf{e}_u)$ with softmax output and trainable parameters $\theta$.
To train it we use the set of training streams represented as tuples $(\mathbf{M}_a, \mathbf{e}_u, c)$
We train it by minimizing the categorical cross-entropy $\mathcal{L}(\mathbf{\hat{c}}_{a,u},\mathbf{c},\mathbf{\theta})$ where $\mathbf{\hat{c}}_{a,u}$ is the estimated probability and $\mathbf{c}$ the one-hot-encoded ground-truth.

\textbf{Practical implementation.}
%
%
The \textbf{audio input}, $\mathbf{M}_a$, is passed to a batch normalization layer then to 4 layers each made of a convolutional (CNN) and a Max-pooling operation. 
The CNNs have various numbers of filters (32, 64, 128, 256) but each with the same size (3x3). They are each followed by a ReLU. 
All Max-poolings are (2x2). 
The flattened output of the last CNN layer is passed to a \ac{fc} layer with 256 units followed by a  ReLU. 
The output of the audio branch $\mathbf{e}_a$ is a 256-d audio embedding vector $\mathbf{e_a}$.
The \textbf{user input}, $\mathbf{e}_u$, is processed through 2 \ac{fc} layers each with a ReLU.
This output is then concatenated with $\mathbf{e}_a$ and passed to a \ac{fc} layer with ReLU activation, and a dropout (with 0.3 ratio) for regularization. 
The final layer is made of $C$ output units with a Softmax activation function, where $C$ is the number of situations to be predicted.
We train the model until convergence by minimizing the categorical cross entropy, optimized with Adam \cite{kingma2014adam} and a learning rate initialized to 0.1 with an exponential decay every 1000 iterations.

\subsection{\acf{sp}}
\label{sitDetails}

The \ac{sp} estimates the probability of each situation $c \in \{1 \ldots C\} $ given a pair (device-data $d$ represented by $\mathbf{d}_{u}^{(t)}$, user $u$ represented by $\mathbf{g}_u$): $P(c | \mathbf{d}_{u}^{(t)}, \mathbf{g}_u)$.
It is implemented as a function $f_{\gamma}(\mathbf{d}_{u}^{(t)}, \mathbf{g_u})$ with Softmax output and trainable parameters $\gamma$.
To train it we use the set of training streams represented as tuples $(\mathbf{g}_u, \mathbf{d}_u^{(t)}, c)$

\textbf{Practical implementation.}
In choosing a real-time ``light'' \ac{sp} model, we prioritize the computational complexity requirements over accuracy. 
The low dimensional input features (11-d = 8 device features + 3 demographic features) already provide a strong case for the investigated models. 
For our implementation, we experimented with different classifiers: Decision Trees, K-Nearest Neighbors, and eXtreme Gradient Boosting (XGBoost) \cite{chen2016xgboost}. 
While all gave comparable results, we chose XGBoost for its consistent performance across splits and different evaluation scenarios. 
Similar to the autotagger model, the output predictions depend on the number $C$ of situations in the dataset.

\section{Evaluation}

We evaluate here the performance of our system which aims at proposing for a given user $u$ a session (sequence of audio-tracks $a$) that fits their current situation $c$.
For this, we first evaluate the performance of the two branches of our system (\ac{uamat} and \ac{sp}) to correctly estimate the situation $c$.
We evaluate this using various numbers of situations: $C \in \{4, 8, 12\}$.
To evaluate the \ac{uamat}, we use the common AUC (Area Under Roc Curve) and Accuracy performance measures.
To evaluate the \ac{sp}, we use the Accuracy but also the Accuracy@$K$. 
This measures the capability of \ac{sp} to include the correct situation in the top $K$ predictions.
We then evaluate the global system by measuring the overlap of correct predictions between the \ac{uamat} and \ac{sp} branches.
This accuracy can be interpreted as the ratio of existing streams that would have occurred in these sessions if the playlists were generated with this system instead. 

\subsection{Scenario}
We approach the evaluation of this system from two different perspectives:
1) evaluating the system on its capability of learning and generalizing,
2) evaluating the proposed system in a stable use-case with frequent users/tracks.

We simulate these scenarios through a different split criteria for the test-set. 
Let the full set of streams in our collected dataset $S$, where each stream $s$ has a user $u$ and a track $a$. 
We will be referring to the training-set as $S_{train}$, the test-set as $S_{test}$, the set of unique users in training and testing as $U_{train}$ and $U_{test}$ respectively, and similarly the unique audio-tracks in the splits as $A_{train}$ and $A_{test}$.

To evaluate the system intelligence and fit to the data, we restrict the evaluation splits to include either: 
1) new users (\textcolor[HTML]{0000ff}{cold-user case}): $S_{test} = \{s | u \notin U_{train}, a \in A_{train}\}$,
2) new tracks (\textcolor[HTML]{8000ff}{cold-track case}): $S_{test} = \{s | u \in U_{train}, a \notin A_{train}\}$. We exclude the specific case of both new tracks and new users because splitting the data with only new user/track pairs in the testset is difficult and rare to find. Additionally, recommending a new track to a new user is not a common nor practical scenario to use for evaluating a system. 

To evaluate how the system would perform in a regular use-case (\textcolor[HTML]{ff0000}{warm case}): $S_{test} = \{s | u \in U_{train}, a \in A_{train}, s \notin S_{train}\}$. 
The regular use-case does not restrict the system to neither new users nor tracks. 
However, the test-set contains exclusively new streams, i.e. (user/track) pairs, not present in the training-set. 
The evaluation of this regular use-case is relatively complex and includes several entwined evaluation criteria. 
The goal is to compare the overlap of generated sessions with groundtruth sessions.

\subsubsection{\ac{uamat} Evaluation}
The results for the \ac{uamat} can be found in Table~\ref{tab:autoTag}. 

As shown, the model can reach satisfying performance relative to the evaluation scenario. 
In terms of AUC, the model's fit for both new users and tracks in the cold user/track splits is not significantly impaired compared to the warm case. 
The performance decreases evidently as the problem gets harder with more situations $C$ to tag, though in some cases it increases given the increase of dataset size from additional situations. 
In terms of accuracy, the model's performance in the intended use-case, i.e. warm case, is satisfying (Accuracy above 70). 
That is to say, the system can correctly tag around two thirds of the user/track listen streams with their correct situational use, when it has seen the user or the track before, but not jointly. 

Note that this accuracy was computed by selecting the most probable situation from the predictions. 
While the high values of AUC (above 0.94) suggest a threshold optimization is needed for each class, in real use-case we do not necessarily need a threshold. 
The prediction probability could be used directly to retrieve tracks, e.g. by ranking tracks with the prediction probabilities and include top ranked tracks in the generated sessions. 
However, this max-probability threshold is needed for further evaluations with the situation predictor and with the sequential retrieval model.

\begin{figure}[h]
    \centering
    \includegraphics[width=\linewidth]{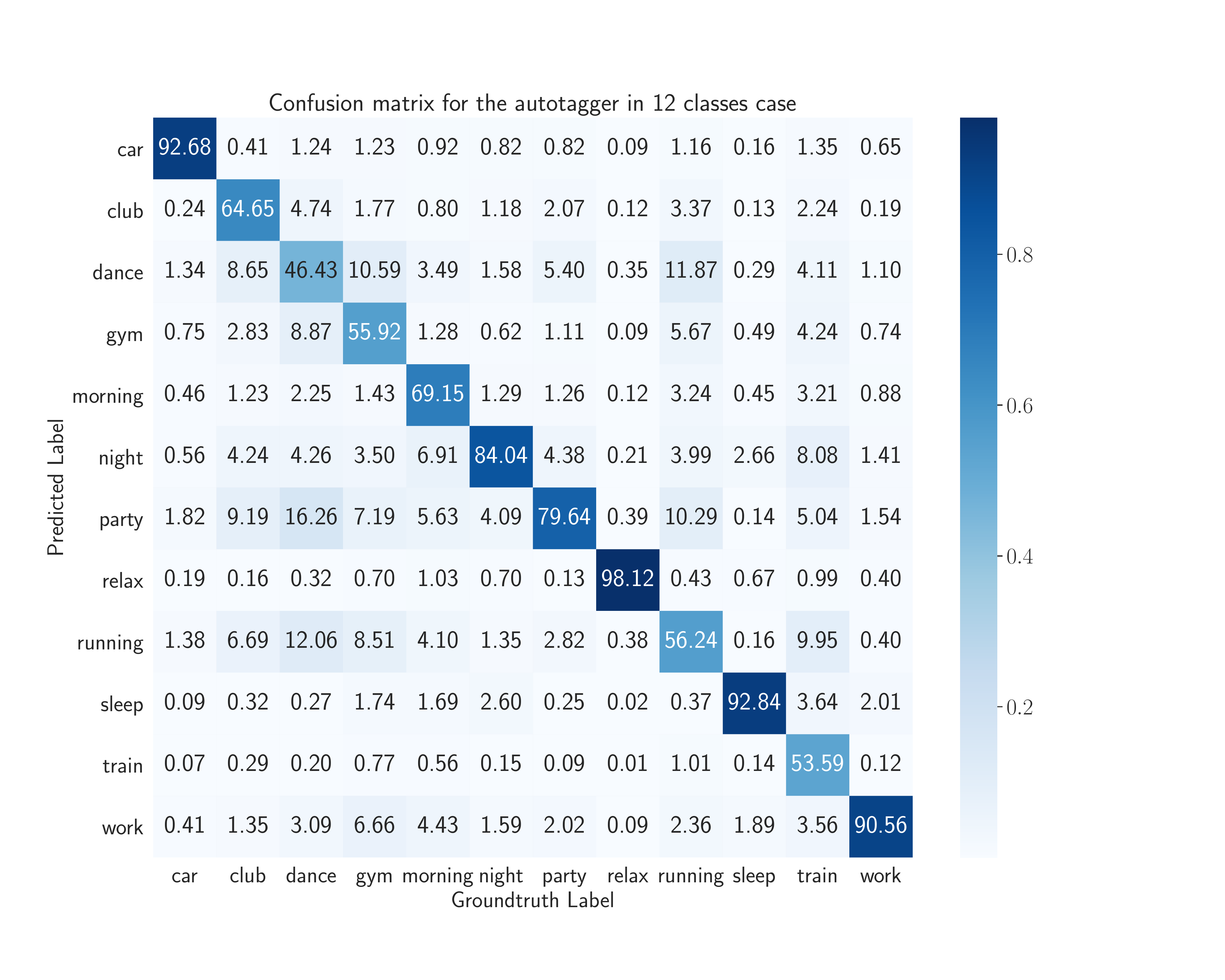}
    \caption{Confusion Matrix of the \ac{uamat} in the case $C$=12 and warm case}
    \label{fig_cm_uamat}
\end{figure}

Additionally, Figure~\ref{fig_cm_uamat} represents the confusion matrix obtained in the $C=12$ and warm case.

\begin{table}[h]
\centering
\caption{Results of the \ac{uamat} evaluated with AUC and Accuracy in the three evaluation protocol splits (cold-user, cold-track, and warm case) and the three subsets of situations (4, 8, and 12). The results are shown as mean(std.).}
\label{tab:autoTag}
\begin{tabular}{c|ccc|}
\hline
\multirow{2}{*}{$C$} & \multicolumn{3}{c}{AUC}   \\ \cline{2-4} 
                                   &\textcolor[HTML]{0000ff}{Cold User} &
                                   \textcolor[HTML]{8000ff}{Cold Track} &
                                   \textcolor[HTML]{ff0000}{Warm} \\ \hline
\cellcolor[HTML]{efefef}4                                  & \cellcolor[HTML]{e6e6ff}0.889 (.009)     & \cellcolor[HTML]{f2e6ff} 0.873 (.013)    & \cellcolor[HTML]{ffe6e6} 0.959 (.013)    \\ 

\cellcolor[HTML]{cccccc}8                                  & \cellcolor[HTML]{ccccff}0.815 (.005)      & \cellcolor[HTML]{e7ccff}0.866 (.007)     & \cellcolor[HTML]{ffcccc}0.945 (.007)    \\ 

\cellcolor[HTML]{999999}12                                 & \cellcolor[HTML]{b3b3ff}0.852 (.004)      & \cellcolor[HTML]{dbb3ff}0.824 (.012)      & \cellcolor[HTML]{ffb3b3}0.941 (.012)     \\ \hline
\multirow{2}{*}{$C$} & \multicolumn{3}{c}{Accuracy}       \\ \cline{2-4} 
                                   &\textcolor[HTML]{0000ff}{Cold User} &
                                   \textcolor[HTML]{8000ff}{Cold Track} &
                                   \textcolor[HTML]{ff0000}{Warm} \\ \hline
\cellcolor[HTML]{efefef}4          & \cellcolor[HTML]{e6e6ff}69.72 (1.07)    & \cellcolor[HTML]{f2e6ff}63.77 (2.33)    & \cellcolor[HTML]{ffe6e6}83.75 (2.33)  \\ 
\cellcolor[HTML]{cccccc}8       & \cellcolor[HTML]{ccccff}47.56 (0.53)    & \cellcolor[HTML]{e7ccff}52.44 (2.31)   & \cellcolor[HTML]{ffcccc}70.81 (1.45)   \\ 
\cellcolor[HTML]{999999}12                & \cellcolor[HTML]{b3b3ff}52.68 (1.25)    & \cellcolor[HTML]{dbb3ff}37.61 (3.47)   & \cellcolor[HTML]{ffb3b3}69.14 (3.79)    \\ \hline
\end{tabular}
\end{table}

\begin{table}[ht]
\centering
\small
\caption{Results of the \ac{sp} evaluated with Accuracy and Accuracy@3 in the three evaluation protocol splits (cold-user, cold-track, and warm case) and the three subsets of situations (4, 8, and 12). The results are shown as mean(std.).}
 \label{tab:SitPred}
\begin{tabular}{c|cc|cc}
\hline
\multirow{2}{*}{$C$} & \multicolumn{2}{c}{Accuracy}       & \multicolumn{2}{c}{Accuracy @3}    \\ \cline{2-5} 
                                   & \textcolor[HTML]{0000ff}{Cold User} & \textcolor[HTML]{ff0000}{Warm} & \textcolor[HTML]{0000ff}{Cold User} & \textcolor[HTML]{ff0000}{Warm} \\ \hline
\cellcolor[HTML]{efefef}4                                   
    & \cellcolor[HTML]{e6e6ff}47.46 (0.98)     &  \cellcolor[HTML]{ffe6e6}66.96 (0.39)      & \cellcolor[HTML]{e6e6ff}90.51 (0.31)          & \cellcolor[HTML]{ffe6e6}96.3 (0.1)     \\ 
\cellcolor[HTML]{cccccc}8                                 
    & \cellcolor[HTML]{ccccff}30.95 (0.89)    & \cellcolor[HTML]{ffcccc}49.23 (0.16)      & \cellcolor[HTML]{ccccff} 64.11 (1.42)    & \cellcolor[HTML]{ffcccc}79.62 (0.13)       \\ 
\cellcolor[HTML]{999999}12                                 
    & \cellcolor[HTML]{b3b3ff}25.00 (0.29)     & \cellcolor[HTML]{ffb3b3}39.92 (0.13)     & \cellcolor[HTML]{b3b3ff}52.04 (0.61)        & \cellcolor[HTML]{ffb3b3}67.62 (0.21)   \\ \hline
\end{tabular}
\end{table}

\begin{figure}[h]
    \centering
    \includegraphics[width=\linewidth]{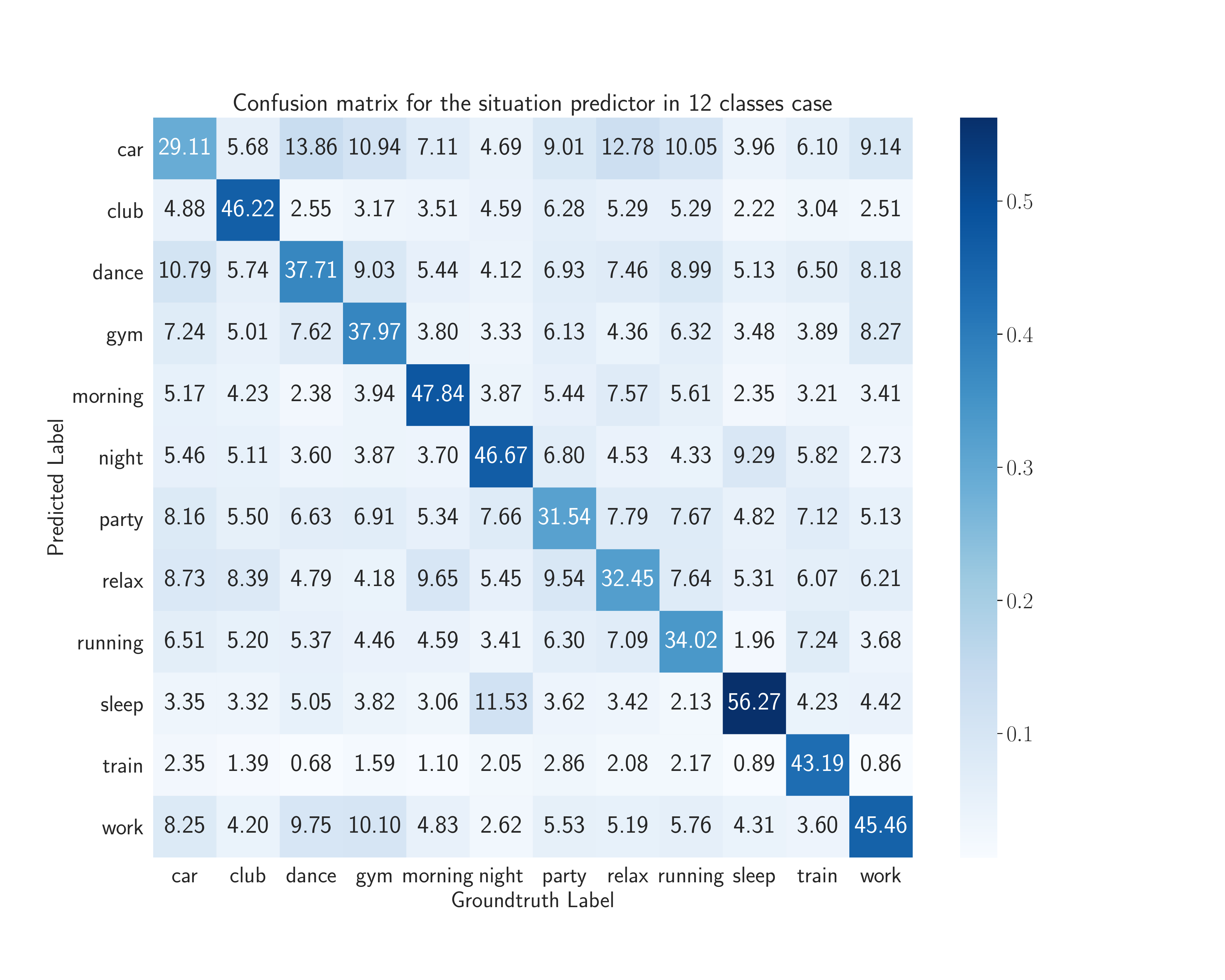}
    \caption{Confusion Matrix of the \ac{sp} with $C$=12 and warm case}
    \label{fig_cm_sp}
\end{figure}

\subsubsection{\ac{sp} Evaluation}
The results for the \ac{sp} can be found in Table~\ref{tab:SitPred}. 
We find that predicting the situation for new users becomes noticeably harder. 
In the case of $C$=12 situations, the system was able to correctly predict the situation for only 25\% of the streams. 
However, when the system is allowed to make multiple guesses (Accuracy@3), the accuracy evidently increases. 
In the case where the user is to make the last decision, the system is able to include the correct situation in the top 3 suggestions 96\%, 80\%, and 68\% in the cases of $C=$4, 8, and 12  situations respectively. 
The choice of $K$, when evaluating with accuracy@$K$, can be obviously changed, and the performance will increase as $K$ increases. 
We choose to display the results for $K$=3 since 3 is around the number of visible items in the carousels displayed by most streaming services on the suggestions screen on mobile devices. 

Additionally, Figure~\ref{fig_cm_sp} shows the confusion matrix obtained in the $C$=12 situations and warm case. 
We observe that the confusion is mostly coherent with the statistic shown earlier of the distribution of situations with the device data. 
Situations that are likely to originate with similar device data are harder to discriminate than the rest. 
For example, we observe a cluster of night-related situations including night, sleep, and relax situations. 
Similarly, outdoors situation are also often confused together. 
Discriminating those situations is hindered by the limited data available. 
However, the convenience of recommending top $k$ situations provides as easy solution to further discriminate between these similar situations.

Finally, to evaluate the challenge in classifying situations from multiple sources, we compare between the evaluation results in each location (France, Brazil) separately. 
We compare between two different cases: 
1) a model trained globally on the data from both locations but tested locally, 
2) a model trained locally on each location independently and tested on the corresponding location. 
Table~\ref{locs} shows the results for this evaluation setting. 
We find that training the models locally slightly improves the results, but not significantly. 
This suggests that using a single unique model for all locations gives comparable results to using multiple local models. 
We also observe a clear distinction in the accuracy between the two locations, where Brazil scores higher than France in all cases. 
This is due to the larger number of users in our dataset who are in France, i.e. there are more users with more distinct patterns in the France case.

\begin{table}[ht]
\centering
\caption{Evaluation results of the globally and locally trained models for each of the two locations in our dataset, France and Brazil, evaluated with accuracy at each subset of situations in the warm case. The results are shown as mean(std.).}
\small
\label{locs}
\begin{tabular}{c||c|c|c|c}
\hline
                             & \multicolumn{2}{c|}{France}                                             & \multicolumn{2}{c}{Brazil}                                             \\ \cline{2-5} 
\multirow{-2}{*}{$C$} & Global                        & Local                         & Global                        & Local                         \\ \hline \hline
4                            & \cellcolor[HTML]{DAE8FC}53.4 (0.2) & \cellcolor[HTML]{B1CFFB}55.1 (0.2) & \cellcolor[HTML]{DAE8FC}59.2 (0.2) & \cellcolor[HTML]{B1CFFB}61.7 (0.2) \\ \hline
8                            & \cellcolor[HTML]{DAE8FC}35.8 (0.1) & \cellcolor[HTML]{B1CFFB}36.8 (0.1) & \cellcolor[HTML]{DAE8FC}45.9 (0.1) & \cellcolor[HTML]{B1CFFB}48.2 (0.1) \\ \hline
12                           & \cellcolor[HTML]{DAE8FC}27.6 (0.1) & \cellcolor[HTML]{B1CFFB}28.2 (0.1) & \cellcolor[HTML]{DAE8FC}39.6 (0.2) & \cellcolor[HTML]{B1CFFB}41.8 (0.1) \\
\end{tabular}
\end{table}

\subsubsection{Joint Evaluation}
The results for the joint system can be found in Table~\ref{tab:joint}. 
As we can see, each variable in our evaluation influences the performance of the system. 
The most influential parameter is the number $C$ of potential situations. 
As the complexity increases, we find the accuracy of the model decreasing: from 58\% in the case $C$=4 with no new users or tracks to 16\% in the case $C$=12 with cold scenarios.
Additionally, we find the expected variation in performance between the cold cases and the warm case of intended use. 
We observe how the drop in the performance of the \ac{uamat} and \ac{sp}, on new users/tracks, negatively affects the joint system performance. 

However, in the harder evaluation case of generating a situational playlists with only 1 guess allowed out of $C$=12, the proposed system would have been able to include at least a third of the actual listened tracks (31.26\%) in those playlists, while pushing them to the user at the exact listened time.

\begin{table}
\centering
\small
\caption{The joint evaluation results of the \ac{uamat} and \ac{sp} and their overlapping predictions evaluated with Accuracy in the three evaluation protocol splits (cold-user, cold-track, and warm case) and the three subsets of situations (4, 8, and 12). The results are shown as mean(std.).}
\label{tab:joint}
\begin{tabular}{c|ccc}
\hline
Model    & \textcolor[HTML]{0000ff}{Cold Users} & \textcolor[HTML]{8000ff}{Cold Tracks} & \textcolor[HTML]{ff0000}{Warm Case} \\ 
\hline \hline
& \multicolumn{3}{c}{\cellcolor[HTML]{efefef}4 Situations}                        \\ \cline{2-4}
\ac{uamat}     
    & \cellcolor[HTML]{e6e6ff}69.73 (1.07)  & \cellcolor[HTML]{f2e6ff}63.78 (2.33)       & \cellcolor[HTML]{ffe6e6}83.75 (2.33)    \\ 
\ac{sp}    
    & \cellcolor[HTML]{e6e6ff}47.46 (0.98)  & \cellcolor[HTML]{f2e6ff}66.81 (0.35)       & \cellcolor[HTML]{ffe6e6}67.20 (0.26)    \\ \hline
\textbf{Overlap}      
    & \cellcolor[HTML]{e6e6ff}36.22 (1.27)  & \cellcolor[HTML]{f2e6ff}44.60 (1.01)        & \cellcolor[HTML]{ffe6e6}\textbf{58.92} (1.71)    \\ 
\hline \hline
& \multicolumn{3}{c}{\cellcolor[HTML]{cccccc}8 Situations}                         \\ \cline{2-4}
\ac{uamat}     
    & \cellcolor[HTML]{ccccff}47.56 (0.53)       & \cellcolor[HTML]{e6ccff}52.44 (2.31)       & \cellcolor[HTML]{ffcccc}70.81 (1.45)   \\ 
\ac{sp}
    & \cellcolor[HTML]{ccccff}30.95 (0.89)       & \cellcolor[HTML]{e6ccff}49.13 (0.24)       & \cellcolor[HTML]{ffcccc}49.35 (0.19)     \\ \hline
\textbf{Overlap}       
    & \cellcolor[HTML]{ccccff}17.77 (0.49)      & \cellcolor[HTML]{e6ccff}28.94  (1.24)       & \cellcolor[HTML]{ffcccc}39.52 (1.27)     \\ 

\hline \hline
& \multicolumn{3}{c}{\cellcolor[HTML]{999999}12 Situations}                      \\ \cline{2-4} 
\ac{uamat}     
    & \cellcolor[HTML]{b3b3ff}52.68 (1.25)     & \cellcolor[HTML]{d9b3ff}37.61 (3.47)     & \cellcolor[HTML]{ffb3b3}69.14 (3.79)     \\ 
\ac{sp}
    & \cellcolor[HTML]{b3b3ff}25.00 (0.29)       & \cellcolor[HTML]{d9b3ff}39.05 (0.31)       & \cellcolor[HTML]{ffb3b3}39.19 (0.14)     \\ \hline
\textbf{Overlap}       
    & \cellcolor[HTML]{b3b3ff}16.19 (0.32)     & \cellcolor[HTML]{d9b3ff} 18.75 (1.63)     & \cellcolor[HTML]{ffb3b3}31.26 (1.30)   \\ \hline
\end{tabular}
\end{table}

\section{Conclusion}
In this study, we address the problem of the unobserved listening situation which influences the users' preferences. 
We proposed a two-branch framework to predict when a situation is being experienced based on the device data, while simultaneously autotagging the music tracks with their intended listening situation in  a personalized manner. 
Through the proposed approach, users could access a set of predicted potential situations. 
These situations are also associated with a set of tracks ``likely'' to be listened to by the user. This likelihood is estimated using an autotagger trained on predicting the situational use of tracks, given a specific user and his/her listening history. We evaluated each of our system's blocks individually and combined. The evaluation results indicated that the system is capable of learning personalized patterns for users, which can be employed to provide contextual music recommendation.

\clearpage
\bibliography{ISMIR2022_template}

%
%
%
%
%

\end{document}